\documentclass{article}
\usepackage[dvips]{graphicx}
\begin{document}

\begin{center}
Poster presented at the 2K1BC Workshop\\
 {\it Experimental Cosmology at Millimetre Wavelengths}\\
 July 9-12, 2001, Breuil-Cervinia (Italy)
\end{center}
\vskip2mm
\hrule
\vskip1cm

\begin{center}
\huge{\bf Sunyaev-Zel'dovich Effect and Morphology of Galaxy Clusters}
\vskip4mm
\large{\bf R. Piffaretti$^\ast$, Ph. Jetzer$^\dagger$, 
D. Puy$^\ast$, S. Schindler$^\ddagger$}
\vskip4mm
{\small
$^\ast$Paul Scherrer Institute 
\\
and Institute of Theoretical Physics, University of Z\"urich - CH
\vskip1.5mm
$^\dagger$Institute of Theoretical Physics, University of Z\"urich 
\\
and Institute for Theoretical Physics, ETH Z\"urich - CH
\vskip1.5mm
$^\ddagger$Astrophysics Research Institute, Liverpool John Moores University - UK}
\end{center}
\vskip9mm
\noindent
{\bf Abstract.} 
We investigate the influence of the finite extension and the aspherical 
geometry of a galaxy cluster on the estimate of the Hubble constant through 
the Sunyaev-Zel'dovich (SZ) effect. An analysis of a recent {\it Chandra} 
image of the galaxy cluster RBS797 indicates a strong ellipticity and thus
a pronounced aspherical geometry. We estimate the total mass of RBS797 assuming spherical or ellipsoidal geometry and show that in the latter case the mass is about 10-17\% less than the one inferred for a spherical shape.
\vskip4mm
\centerline{\bf \large Introduction}
\vskip2mm
\noindent
The recent technical developments of millimetre receivers open 
new perspectives for more accurate measurements of the SZ effect
and will thus trigger new developments on theoretical work 
(see the invited paper of Y. Rephaeli in this conference). Accurate measurements of the SZ effect are still difficult as well as their
correct interpretation, indeed 
systematic errors can be significant. For example Cooray \cite{Cooray:1998} 
showed the influence of projection effects and Sulkanen \cite{Sulkanen:1999} 
pointed out that the shape of galaxy clusters could produce systematic errors.
More recently, Puy et al. \cite{Puy:2000} 
investigated the SZ effect and the $X$-ray surface brightness for galaxy
clusters with a non-spherical mass distribution. 
\\
In the first part of this communication we review the ``classical'' 
systematic errors such as cluster extension and geometry, by discussing
the recent millimetre 
measurements of Mauskopf et al. \cite{Mauskopf:2000} 
of the SZ effect in Abell 1835. 
In the second part, we briefly comment on the possible geometrical 
influence on the determination of the total mass of galaxy clusters.
\vskip4mm
\centerline{\bf \large SZ effect and the Hubble constant}
\vskip2mm
\noindent
Observations of galaxy clusters in the millimetre and $X$-ray wavebands give 
important information for cosmology. By combining the SZ intensity change and 
the $X$-ray emission observations, the angular diameter distance to galaxy 
clusters can be derived. Assuming a cosmological model, this leads to an 
estimate of the Hubble constant $H_o$.\\
Mauskopf et al. \cite{Mauskopf:2000} 
determined $H_o$ from $X$-ray measurements of A1835 obtained with $ROSAT$ and 
from the corresponding millimetric observations of the SZ effect with the 
{\it Suzie} experiment. Assuming an infinitely extended, spherical gas 
distribution with an isothermal profile $\beta=0.58 \pm 0.02$, 
$T_{eo}=9.8 ^{+2.3}_{-1.3}$ keV, $n_{eo} = 5.64 ^{+1.61}_{-1.02} 
\times 10^{-2}$ cm$^{-3}$, they found $H_o=59^{+36}_{-28}$ km s$^{-1}$ 
Mpc$^{-1}$. 
\\
Since the hot gas in a real cluster has a finite extension, each 
of the observed quantities as the Compton parameter $y$ and the $X$-ray surface 
brightness $S_x$ will be smaller than those estimated based on the 
infinite extension assumption 
($l \rightarrow \infty$). Since the Hubble constant is estimated 
from the ratio $S_x/y^2$, in Puy et al. \cite{Puy:2000} we showed that the 
relative error $\epsilon_{H_0}^{fini}$ on the estimate of the Hubble 
constant, between a spherical distribution with and without finite extension, 
is given by:

\begin{eqnarray}
\epsilon_{H_0}^{fini} &=&
\frac{H_0(\infty) - H_0(l)}{H_0(\infty)}
\nonumber \\
 &=&  1- 
\frac{ 
B(3 \beta-\frac{1}{2},\frac{1}{2})
\left[
B(\frac{3}{2}\beta-\frac{1}{2},\frac{1}{2})-
B_m(\frac{3}{2}\beta-\frac{1}{2},\frac{1}{2}) 
\right]^2
}
{
B^2(\frac{3}{2}\beta-\frac{1}{2},\frac{1}{2})
\left[
B(3 \beta-\frac{1}{2},\frac{1}{2}) - 
B_m(3 \beta-\frac{1}{2},\frac{1}{2})
\right]
}~.
\end{eqnarray}
The functions $B$ and $B_m$ are combinations
of the classical Gamma-functions and 
the factor $m$ is a cut-off relative to the finite extension 
(see Puy et al. \cite{Puy:2000}). In the same way we have analysed the 
relative error between spherical and aspherical geometries 
(without finite extension). 
\\
Figure 1 
shows the influence of geometry and of the assumption of finite extension on 
the above result using the same input parameters of Mauskopf et al. 
\cite{Mauskopf:2000}. The left panel shows that 
for a spherical geometry $H_o$ displays a strong dependence on the cluster 
extension. The right panel gives the value of $H_o$ assuming an infinite 
extended ellipsoid shaped cluster as a function of its axis ratio 
$\zeta_1/\zeta_3$.

\begin{figure}[h]
\begin{center}
 \resizebox{.5\columnwidth}{!}
  {\includegraphics[angle=-90]{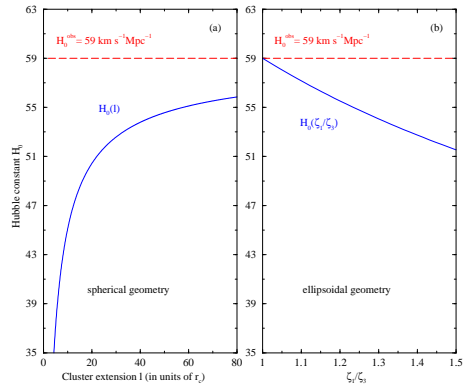}}
\caption{The Hubble constant derived from the data of Mauskopf 
et al. \cite{Mauskopf:2000}. 
Fig. (a) shows the influence of finite extension, 
while Fig. (b) gives the value of $H_o$ assuming an axisymmetric ellipsoidal 
geometry. In the latter case, oblate or prolate geometry give the same 
value of $H_o$ when taking a line of sight through the cluster center, 
as is assumed here.}
\end{center}
\end{figure}

We see that it is crucial to know the shape of a cluster and its temperature 
profile. In this context the $X$-ray satellites $XMM$ and \textit{Chandra} have the 
necessary spatial and spectral resolution to address this problem on 
nearby cluster samples.
\vskip4mm
\centerline{\bf \large Morphology of galaxy clusters}
\vskip2mm
\noindent
The $\beta$-model \cite{Cavaliere:1976} is widely used in $X$-ray astronomy 
to parametrise the gas density profile in clusters of galaxies by fitting 
their surface brightness profile. In this fitting procedure spherical symmetry
 is usually assumed, also in cases where the ellipticity of the surface 
brightness isophotes is manifest. For example Fabricant et al. \cite{Fabricant:1984} showed a pronounced ellipticity of the 
surface brightness for the cluster Abell 2256, Allen et al. \cite{Allen:1993}
 obtained the same result for the profile of Abell 478 and 
Neumann \& B\"ohringer \cite{Neumann:1997} for CL0016+16.\\
The asphericity of the observed surface brightness let us also ponder on the 
possible asphericity of the intracluster medium, 
which can be modelled with an ellipsoidal 
$\beta$-model rather than with the less accurate spherical one.\\
Hughes \& Birkinshaw \cite{Hughes:1998} fitted the surface brightness of 
CL0016+16, which shows an axis ratio of major to minor axis of 1.176, with 
both circular and elliptical isothermal $\beta$-models obtaining for the best 
fit parameters $\beta^{circ}=0.728^{+0.025}_{-0.022}$, 
$\sigma_c^{circ}=0.679^{+0.045}_{-0.039}$ arcmin and 
$\beta^{ell}=0.737^{+0.027}_{-0.022}$, $\sigma_c^{ell}=0.746^{+0.044}_{-0.044}$ arcmin ($\sigma_c^{ell}$ is the core radius along the major axis), respectively, with the latter model providing a 
considerably better fit.
\\
More recently, a {\it Chandra} observation of the galaxy cluster 
RBS797 reveals a pronounced aspherical geometry \cite{Schindler:2001}.
The analysis of the image (see Figure 2) gives a strong ellipticity, where the axis ratio of major to minor axis varies slightly from 1.3 at a radius of 0.26 arcmin to 
1.4 at a radius of 1.7 arcmin (as mentioned in Schindler et al. 
\cite{Schindler:2001}). Our analysis of the surface brightness profile for RBS797 gives best fit parameters: $\beta^{circ}=0.62^{+0.03}_{-0.03}$, $\sigma_c^{circ}=7.32^{+0.7}_{-0.7}$ arcsec ( see also \cite{Schindler:2001}) and $\beta^{ell}=0.59^{+0.02}_{-0.02}$, $\sigma_c^{ell}=7.89^{+0.9}_{-0.9}$ arcsec 
(along the major axis), for the circular and elliptical models respectively \cite{Piffaretti:2001}.
\\
Assuming hydrostatic equilibrium the total mass of a 
cluster can be estimated from the parameters provided by the surface 
brightness fit and clearly some care is needed if 
the galaxy cluster in question shows a 
pronounced ellipsoidal shape. Assuming hydrostatic equilibrium, the general expression for the total mass density $\rho_{tot}$ is given by
\begin{equation}
\rho_{tot}=-\Big(\frac{k_B}{4\pi G \mu m_p}\Big)\vec{\nabla}
\Bigg[\frac{1}{\rho_g}\vec{\nabla}(T_g \rho_g)\Bigg],
\end{equation}
where $\rho_{g}$ is the gas mass density, $T_g$ its 
temperature and $\mu m_p$ is the mean particle mass of the gas.\\
In order to obtain the total mass of the cluster, one assumes a spherical geometry and integrates this equation over a sphere with radius $R$. For RBS797 we have additionally investigated the ellipsoidal geometry (for which we assume oblate or prolate shapes) and integrated the equation over an ellipsoid 
(concentric and similar to the gas core) with major semi-axis $R$ \cite{Piffaretti:2001}. We thus obtained mass estimates for the spherical shape: $M_{tot}^{sph}(R=4 \sigma_c^{circ}=29.28 \, $arcsec$)=8.66^{+2.5}_{-2.3} \times 10^{13} M_{\odot}$ and $M_{tot}^{sph}(R=30 \sigma_c^{circ}=219.6 \, $arcsec$)=6.89^{+2.0}_{-1.8} \times 10^{14} M_{\odot}$ and for ellipsoidal shapes, which are, compared at the same values of $R$, lower than those for the spherical symmetry by $\sim 10 \%$ and $\sim 17 \%$ for oblate or prolate shapes, respectively.  

\begin{figure}[h]
\begin{center}
 \resizebox{.5\columnwidth}{!}
  {\includegraphics{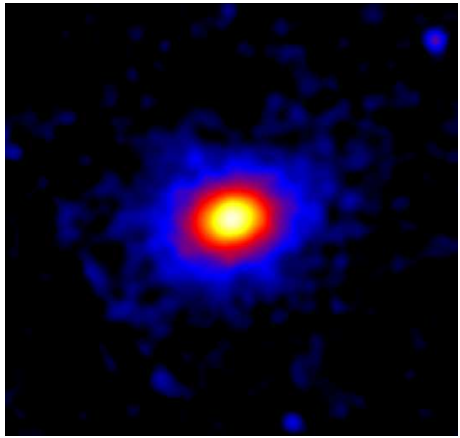}}
\caption{\textit{Chandra} image of the cluster RBS797. The cluster is 
rather regular with, however, an ellipticity of 1.3-1.4 in E-W direction. The 
center and the position angle ($\approx -70^{\circ}$, N over E) of the 
various isophotes are almost the same over the entire radius range, from Schindler et al. 
\cite{Schindler:2001}.}
\end{center}
\end{figure}

\centerline{\bf \large Acknowledgments}
\vskip2mm
\noindent
We would like to thank Marco de Petris, Massimo Gervasi and Fernanda 
Luppinacci for organizing 2K1BC Workshop and making it such an enjoyable 
and stimulating meeting. This work has been supported by the {\it D$^r$ 
Tomalla Foundation} and by the Swiss National Science Foundation.

\end{document}